\documentclass[fleqn,usenatbib]{mnras}

\usepackage{newtxtext,newtxmath}
\usepackage{soul}
\usepackage[T1]{fontenc}
\DeclareRobustCommand{\VAN}[3]{#2}
\let\VANthebibliography\thebibliography
\def\thebibliography{\DeclareRobustCommand{\VAN}[3]{##3}\VANthebibliography}

\usepackage[dvipsnames]{xcolor}

\usepackage{graphicx}
\usepackage{amsmath}
\usepackage{amssymb}
\usepackage{changes}
\usepackage{hyperref}

\usepackage{adjustbox}
\usepackage{anyfontsize}
\usepackage{longtable}
\usepackage{afterpage}

\title[Spatial connection between AGN flares and GWs]
{AGN flares as counterparts to LIGO/Virgo mergers:
No confident causal connection in spatial correlation analysis}

\author[N. Veronesi et al.]{
Niccolò Veronesi,$^{1}$\thanks{E-mail: veronesi@strw.leidenuniv.nl}
Sjoert van Velzen,$^{1}$
Elena Maria Rossi,$^{1}$
\\
$^{1}$ Leiden Observatory, Leiden University, PO Box 9513, 2300 RA Leiden,
The Netherlands}

\date{Accepted 2024 December 17. Received 2024 November 26; in original form 2024 May 8}

\pubyear{2023}

\begin{document}
\label{firstpage}
\pagerange{\pageref{firstpage}--\pageref{lastpage}}
\maketitle

\begin{abstract}
The primary formation channel for the stellar-mass Binary Black Holes
which have been detected merging by the LIGO-Virgo-KAGRA (LVK) collaboration
is yet to be discerned. One of the main reason is that the detection of
an Electromagnetic counterpart to such Gravitational Wave (GW) events,
which could signpost their formation site, has so far been elusive. Recently,
20 Active Galactic Nuclei flaring activities detected by the Zwicky Transient
Facility have been investigated as potential counterparts of GW events by
Graham et al. (2023). We present the results of a spatial correlation analysis
that involves such events and uses the up-to-date posterior samples of 78
mergers, detected during the third observing run of the LVK collaboration.
We apply a likelihood method which takes into account the exact position of
the flares within the 3D sky map of the GW events. We place an upper limit
of 0.155 at a 90 per cent credibility level on the fraction of the detected
coalescences that are physically related to an observed flare, whose posterior
probability distribution peaks at a null value. Finally, we show that the
typically larger values of the masses of the GW-events, which host at least
one flare in their localisation volume, are also consistent with the
no-connection hypothesis. This is because of a positive correlation between
binary masses and the size of the localisation uncertainties.
\end{abstract}



\section{Introduction}
\label{sec:intro}

During the third observing run (O3) of the LIGO-Virgo-KAGRA (LVK)
collaboration, Gravitational Waves (GWs) coming from 80 confirmed
mergers of binaries of compact objects were detected
\citep{Abbott+21,Abbott+20d}. Several channels for the formation
of the merging systems have been proposed
\citep[see ][for a recent review]{Mapelli21}. These pathways can
be divided into two main categories: the evolution of isolated
stellar binary systems, and the formation inside dense environments,
such as Nuclear Star Clusters or accretion discs of Active Galactic
Nuclei (AGN). The formation of merging Binary Black Holes (BBHs)
in AGN discs is expected to be facilitated not only by the high
density of compact objects, but also by their dynamical interaction
with the gas
\citep{Stone+17,Li_Lai22,Li_Lai23,Li_Lai23b,Qian+23,Rowan+23,Rowan+23b,RodriguezRamirez24a}.
For this reason, the so-called ‘‘AGN-channel’’ for the formation
of merging binaries has been recently studied and modelled
extensively.

The component masses of binaries coalescing in accretion discs
around Massive Black Holes are expected to populate the high
end of the astrophysical stellar-mass Black Hole mass spectrum
presented in \citet{Abbott+23a}. This is mainly because Type
I migration caused by the interaction between a gaseous disk and
the compact objects within it makes mass segregation very
efficient, since the migration time-scale is inversely proportional
to the mass of the compact object \citep{Armitage07,McKernan+11,Secunda+19}.
Migration can therefore increase the density of compact
objects in the inner region of the AGN disc, where binary formation
and hardening can be efficiently assisted by the the interaction
with the gaseous environment \citep{Tagawa+20,Yaping21,Yaping22,DeLaurentis+23},
unless it presents an elevated level of turbulence \citep{Wu24}.

The vicinity to the central Massive Black Hole
implies that mergers of compact objects inside an AGN accretion
disc happen in the presence of a deep gravitational potential.
For this reason, the speed at which the remnant objects get
kicked from the location of the merger, typically of the order
of hundreds of kilometres per second, is likely lower than the
escape speed of the environment. Therefore there is the possibility
for it to be retained and to take part to another BBH merger.
In this scenario of hierarchical mergers, the components of the
binaries are expected to have high masses and dimensionless spins
close to $\approx0.7$, which is the value that the remnant of a
previous merger is expected to have, due to total angular
momentum conservation \citep{Gerosa_Berti17,Gerosa_Fishbach21}.
In addition, binaries of compact objects merging inside an AGN
disc have been modelled as able to develop and maintain a
measurable level of eccentricity also in the LVK frequency range
\citep{Samsing+22,Calcino+23,Fabj_Samsing24}. The detection of
non-zero eccentricity in one or more GW signals would therefore
represent a strong hint regarding the efficiency of disc-assisted
BBH formation.
Finally, the symmetry of an astrophysical environment such as
a rotating disc has been claimed to be sufficient to explain
the anti-correlation between the binary effective spin parameter
$\chi_{\rm eff}$ and the mass ratio $q$ that has been observed
in the population of mergers detected by the LVK collaboration
\citep{McKernan+21,Santini+23}.
The comparison of the binary parameters' distributions predicted
for the AGN channel with the ones of detected GW events is a
viable approach for estimating the fractional contribution of
this formation path to the total BBH merger rate. For example,
with this method \citet{Gayathri23} estimated a fractional
contribution to the astrophysical BBH merger rate of approximately
20 per cent.

Alternatively, it is possible to put constraints on the
the AGN channel's efficiency
by analysing the spatial correlation between the sky maps
of the events detected by the LVK interferometers and the
positions of observed AGN \citep{Bartos+17,Veronesi+22,Veronesi+23}.
In particular, in \citet{Veronesi+23} was found that the
fraction of detected BBH mergers that took place in an AGN
brighter than $10^{45.5}{\rm erg\ s}^{-1}$
($10^{46}{\rm erg\ s}^{-1}$) is lower than $0.49$ ($0.17$)
at 95 per cent credibility level. This result tends to agree
with the conclusions of the theoretical work presented in
\citet{Grishin24}, where it is suggested that a reduced efficiency
in creating over-densities of compact objects inside discs of
luminous AGN is to be expected when taking into account prescriptions
for Type I migration that are calibrated from 3D simulations
\citep{Jimenez17}.

A third way to identify the host environment of the merging
binaries is the direct detection of a transient Electromagnetic
(EM) counterpart. The vast majority of the detected GW events
are coalescences of BBHs, which are not generally expected to
produce a detectable associated EM signal. However, such
counterparts are expected to be produced, even if not necessarily
detectable, whenever mergers take place in gaseous environments,
like AGN accretion discs \citep{Bartos+16,McKernan+19}.
A possible origin of these counterparts is Bondi accretion on the
merging objects at a hyper-Eddington rate, that triggers a
Bondi explosion \citep{Wang+21b}. Another potential source of a
detectable EM transient is a jet coming from the accretion
of magnetized medium onto the remnant object that traverses the
gaseous disc after receiving a recoil kick from the merger event
\citep{Chen_Dai23}. Recently, \citet[][G23, hereafter]{Graham+23} 
have identified 20 unusual AGN flaring activities observed by
the Zwicky Transient Facility \citep[ZTF;][]{Bellm+19,Graham+19}
as potential EM counterparts of the GW events detected during O3.
These transients have been labelled as not caused by Supernovae
(SNe), Tidal Disruption Events (TDEs), or regular AGN variability.
A possible physical cause of such flares is Bondi drag accretion
and shock of the gas that interacts gravitationally with the
kicked remnant of a merger event that happened inside the disc.
The flare is expected to be manifest when such remnant reaches the
$\tau=1$ optical depth surface of the disc. In \citet{Tagawa+24}
is presented an emission model based on the presence of a Blandford-Znajek
jet produced from BHs in AGN discs \citep{Tagawa+22,Tagawa+23b},
and it is applied to the flares reported in G23, finding that such
model can be consistent with the observed events after a number of
assumptions were made, mainly regarding the accretion rate onto
the stellar-mass BH, and the fraction of jet power that ends up
in radiation.

In this work we present the results of a spatial and temporal
correlation study for the identification of the host environment
of the mergers detected by the LVK collaboration. In particular,
this analysis focuses on the connection between the mergers
detected during O3 and the 20 EM transients presented and
examined in G23. 

We calculate the posterior probability distribution of
$f_{\rm flare}$, the fraction of the LVK events detected during
O3 that are causally connected to an AGN flare. We do this by
using an adapted version of the statistical approach presented
in \citet{Veronesi+23}. This is a model-independent approach, we
only take into account spatial and temporal correlation, but
otherwise remain agnostic about which AGN and GW events are
favorable counterparts. 

In the past, Bayesian analyses have already been performed
to investigate the spatial correlation between GW events and AGN
flaring activities. In particular, the relation between the BBH
merger GW190521 and a flare of the AGN J124942.3+344929,
that had been proposed as a candidate EM counterpart by \citet{Graham20},
has been investigated by \citet{Palmese21}. Their spatial correlation
analysis concluded that the match between the GW event and the EM
transient has a probability of approximately 70 per cent of being
caused by chance coincidence. The lack of a confident physical
association between the same two events is also found by
\citet{Ashton21}.

In Section \ref{sec:dataset} we present the properties of the data
that have been used in the analysis, which include the GW detections
and the  AGN flares considered as potential EM counterparts. The
description of the statistical method used to check the
significance of the GW-AGN flares connection is in Section
\ref{sec:method}, while the results of its application to the observed
data are presented in Section \ref{sec:res}. Finally, in Section
\ref{sec:concl} we draw final conclusions from our results and
discuss what are their implications concerning the physical relation
between BBH mergers and AGN flares. 

We adopt the cosmological parameters of the Cosmic Microwave
Background observations by Planck \citep{Planck+15}:
$H_0=(67.8\pm0.9)\ {\rm km}\ {\rm s}^{-1}{\rm Mpc}^{-1}$,
$\Omega_{\rm m}=0.308\pm0.012$, and $n_{\rm s}=0.968\pm0.006$.


\section{Datasets}
\label{sec:dataset}

In this section we first present the main properties of the AGN
flares selected in G23 as potential EM counterparts to O3
detections. We then describe and list the GW events we use
in our analysis. Finally we show all the spatial and temporal
associations that exist between the two catalogues, both in the
case of a 3D cross-match, and in the case of a 2D sky-projected one.

\subsection{AGN flares}
\label{subsec:flares}
All the AGN flares we use in our analysis have been observed
by ZTF. This facility uses a 47${\rm deg}^2$ field-of-view
camera to cover the majority of the sky above a declination of
$-30^\circ$ every two or three nights in the g-band and in
the r-band. To select potential EM counterparts for the GW
events detected during O3, in G23 the data from the fifth
release (DR5) \footnote{https://www.ztf.caltech.edu/page/dr5}
were used, and the search has been limited to AGN at a
redshift $z\leq1.2$, given the sensitivity the interferometers
of LIGO and Virgo had during O3. The light curves of the
observed AGN flares have been fitted with a Gaussian rise
- exponential decay form (see Equation 13 in G23). From the
original full sample were then removed all the flares that
have been considered as originated from SNe, TDEs, or
regular AGN variability. This selection has been done based
on the timescales of the transients, their g-r colour,
their rate of colour evolution, and their total observed
energy. The resulting selected sample consists of 20 AGN
flares that are considered as potential EM counterparts for
the GW events detected during O3. All these flares have
peaked during the time window that can allow a causal
connection with at least one merger. This temporal match
is considered possible if the AGN has not flared before
the first GW detection or more than 200 days after the
last one. Any merger happening after that time is not
considered as a potential counterpart, since the estimated
time required for the remnant of the merger to reach
the edge of the accretion disc is of the order of tens
of days.

\begin{figure}
    \centering
    \includegraphics[trim= 110 0 0 0,clip,width=1.05\columnwidth]{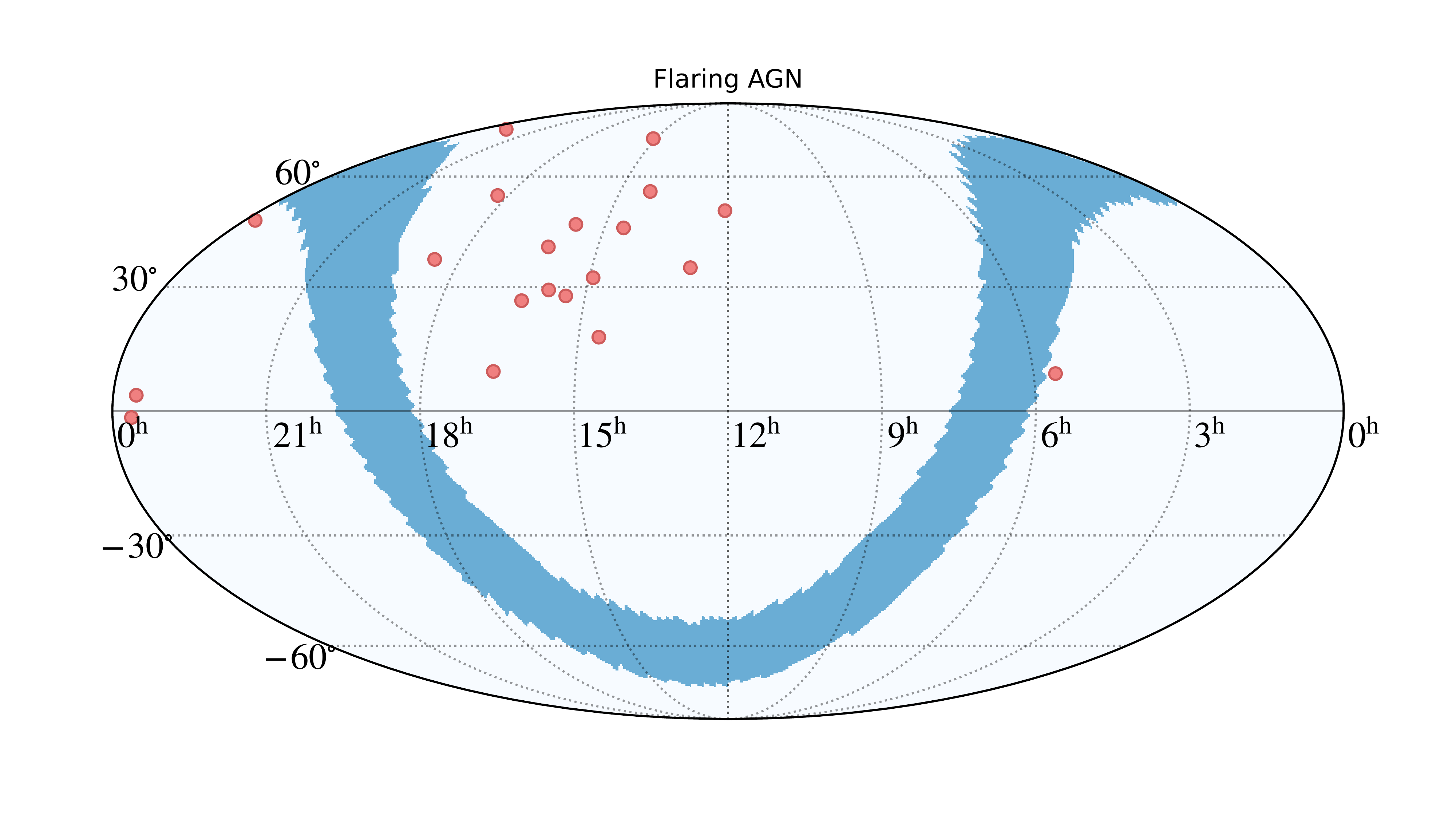}
    \caption{Mollweide projection of the sky position of the 20 AGN
    flare identified as potential EM counterparts of GW events in G23.
    The resolution corresponds to the one of an HealPix map with
    NSIDE=32. The pink markers correspond to the position of the
    flares, while the light blue region correspond to pixels that
    have a galactic latitude between $-10^{\circ}$ and $10^{\circ}$.
    They correspond to the region that includes the galactic plane.}
    \label{fig:skyflares}
\end{figure}

In Table \ref{tab:flares} we list the main properties
of the 20 potential counterparts. In particular we report
the name of the AGN, the Right Ascension (RA) and Declination
(Dec), the redshift, the Modified Julian Day (MJD) of the
peak, and the Gaussian rise time of the light curve, $t_g$.
The sky distribution of these transients is shown in
Figure \ref{fig:skyflares}.
\begin{table*}
    \caption{List of the 20 AGN flares that were selected
    in G23 as potential EM counterparts of GW events
    detected during O3. We report the name of the flaring
    object, its Right Ascension and Declination, its
    redshift, the MJD of the peak, and the Gaussian rise
    timescale. For J053408.41+085450.6 we use the same
    photometric estimate of the redshift that has been
    used in G23. For the AGN located at J150748.68+723506.1
    and at J234420.76+471828.9 there is no available
    redshift estimate in the literature. These two
    sources are therefore excluded from the three-dimensional
    spatial correlation analysis presented in the following
    sections.}
    \begin{tabular}{|c|c|c|c|c|c|}
    \hline
    AGN name & RA & Dec & Redshift & ${\rm MJD}_{\rm peak}$ & $t_{\rm g}$ \\
     & [deg] & [deg] & & & [days] \\
    \hline
    J053408.42+085450.7 & 83.535 & 8.914 & 0.5* & 58890 & 17 \\
    J120437.98+500024.0 & 181.158 & 50.007 & 0.389 & 58894 & 17 \\
    J124942.30+344928.9 & 192.426 & 34.825 & 0.438 & 58671 & 16 \\
    J140941.88+552928.1 & 212.425 & 55.491 & 0.074 & 58616 & 11 \\
    J143157.51+451544.0 & 217.990 & 45.262 & 0.693 & 58859 & 12 \\
    J143536.15+173755.4 & 218.901 & 17.362 & 0.095 & 58673 & 9 \\
    J145500.22+321637.1 & 223.751 & 32.277 & 0.177 & 58590 & 14 \\
    J150748.68+723506.1 & 226.953 & 72.585 & --- & 58971 & 20 \\
    J152433.35+274311.6 & 231.139 & 27.720 & 0.069 & 58611 & 8 \\
    J154342.46+461233.4 & 235.927 & 46.209 & 0.599 & 58975 & 21 \\
    J154806.31+291216.3 & 237.026 & 29.205 & 1.090 & 58997 & 55 \\ 
    J160822.16+401217.8 & 242.092 & 40.205 & 0.627 & 58921 & 27 \\
    J161833.77+263226.0 & 244.641 & 26.541 & 0.126 & 58600 & 11 \\
    J163641.61+092459.2 & 249.173 & 9.416 & 1.155 & 58694 & 9 \\
    J181719.95+541910.0 & 274.333 & 54.319 & 0.234 & 58784 & 17 \\
    J183412.42+365655.2 & 278.552 & 36.949 & 0.419 & 58689 & 13 \\ 
    J224333.95+760619.2 & 340.891 & 76.105 & 0.353 & 58772 & 11 \\
    J233252.05+034559.7 & 353.217 & 3.767 & 1.119 & 58855 & 40 \\
    J233746.08-013116.3 & 354.442 & -1.521 & 0.115 & 58703 & 8 \\
    J234420.76+471828.9 & 356.087 & 47.308 & --- & 58669 & 11 \\
    \hline
    \end{tabular}
    \label{tab:flares}
    \end{table*}


\subsection{GW events}
\label{subsec:gws}

We make use of 78 GW events detected during O3, which are contained in
the Gravitational Wave Transient Catalog (GWTC) 2.1 \citep{Abbott+20d}
and in GWTC 3 \citep{Abbott+21}. The former includes all the events
detected up to the first half of O3, O3a, started on April 1st, 2019
and ended on October 1st of the same year, while the latter includes
all the events detected during the second half of O3, O3b, that
started on November 1st, 2019, and ended on March 27th, 2020. We
only had to exclude GW200308\_173609 and GW200322\_091133 because
for these two poorly-localized events the available data do not
allow the evaluation of the size of the 90 per cent Credibility Level
localisation volume (V90).

We work with the GW sky maps obtained from the latest (at the
time in which this paper is written) versions of the posterior samples
published by the Gravitational Wave Open Science Center (GWOSC)
\citep{Abbott+23}. Such samples
were published on May 13th, 2022 for GWTC-2.1, and on June 23rd,
2023 for GWTC-3. Since G23 has been published in January 2023,
we make use of a different, updated version of the posterior
distributions for the events of GWTC-3, with respect to such work.
The same is true for all the GW events for which in G23 GWTC-2.0
has been used. For such mergers, we use the posterior
distributions of GWTC-2.1. The posterior samples coming from
the {\scshape IMRPhenomXPHM} \citep{Pratten+21} waveform model
have been used for every GW event a part from GW190425\_081805,
GW191219\_163120, GW200105\_162426, and GW200115\_042309, for
which the {\scshape Mixed} posterior samples were used.

To reproduce the analysis of G23, we have to estimate the fraction
of the 90 per cent credibility level localisation area (A90) that is
outside the galactic plane and has been observed by ZTF at least 20
times in both the g-band and r-band in the 200 days following
the LVK detection. This threshold number of observations is used to be
consistent with G23. We call this parameter $f_{\rm cover}$. To
calculate it we use the ZTF forced photometry service
\citep{Masci+23} to obtain the number and the MJD of the
observations at each sky location. The values of $f_{\rm cover}$
for all the sky maps we use in this work are listed in Table
\ref{tab:gws}, together with the ID of the corresponding GW event,
the catalogue it is contained in, the sizes of A90 and V90, and
the MJD of the detection. The sizes of V90 we report refer to
comoving localisation volumes, and not to Euclidean volumes in
luminosity distance. This choice implies a cosmological model to be
assumed during the cross-matching process, but leads to no difference
in its results.


\subsection{Matching events}
The positions of the 20 flares listed in Table \ref{tab:flares}
are cross-matched with the sky maps of the GW events listed in
Table \ref{tab:gws}. This is done using the {\tt postprocess.crossmatch}
function of the {\tt ligo.skymap} package, searching space by
descending probability density per unit comoving volume. The
cross-matching is performed both in the 3D comoving space, and in
the 2D sky-projected space. For the 3D case, the two AGN flares
without any redshift information are excluded. This is consistent with
the approach used in G23, since these two potential counterparts
are also not contained in their sample of 7 matching events. 

To be considered as matching with a merger event, an AGN flare
has to be found within the 90 per cent Credibility Level
localisation volume (or area) of the sky map. Following G23, we
also require that the difference between the flare's peak time
and the GW detection time is smaller than 200 days and larger
than the flare's Gaussian rise time.

The 3D (2D) spatial and temporal cross-match of the 20 flares
with the sky maps of the mergers detected during O3 finds that 11
(19) GW events and 8 (17) AGN flares match to yield a total of 12
(32) one-to-one matches. In other words, we find that for 12 (32) times
one of the 8 (17) flares has peaked in one of the 11 (19) GW localisation
volumes during the temporal window allowed for the match.

The one-to-one matches are listed in Table \ref{tab:matches1} and
Table \ref{tab:matches2}. For each of them we list the ID of the
corresponding GW event, the name of the flaring AGN, and the credibility
level corresponding to its position within the sky map of the merger.

\begin{table}
    \caption{List of the 3D matches between GW events detected
    during O3 and AGN flares contained in ZTF DR5 and identified
    as potential EM counterparts. We report the ID of the GW event,
    the name of the flaring AGN, and the credibility level of its
    position within the localisation volume of the binary merger.}
    \begin{tabular}{|c|c|c|}
    \hline
    GW ID & AGN name & Credibility level\\
    \hline
        GW190514\_065416 & J224333.95+760619.2 & 0.175 \\
        GW190521\_030229 & J124942.30+344928.9 & 0.360 \\
        GW190620\_030421 & J124942.30+344928.9 & 0.778 \\
        GW190706\_222641 & J183412.42+365655.2 & 0.181 \\
        GW190708\_232457 & J233746.08-013116.3 & 0.839 \\
        GW190719\_215514 & J181719.95+541910.0 & 0.185 \\
        GW190731\_140936 & J053408.42+085450.7 & 0.479 \\
        GW190803\_022701 & J053408.42+085450.7 & 0.644 \\
        GW190803\_022701 & J120437.98+500024.0 & 0.543 \\
        GW200128\_022011 & J120437.98+500024.0 & 0.879 \\
        GW200216\_220804 & J154342.46+461233.4 & 0.727 \\
        GW200220\_124850 & J154342.46+461233.4 & 0.137 \\
    \hline
    \end{tabular}
    \label{tab:matches1}
    \end{table}

\begin{table}
    \caption{List of the 2D sky-projected matches between GW
    events detected during O3 and AGN flares contained in ZTF
    DR5 and identified as potential EM counterparts. We
    report the ID of the GW event, the position of the flaring
    AGN name, and the credibility level of such position within
    the localisation area of the binary merger.}
    \begin{tabular}{|c|c|c|}
    \hline
    GW ID & AGN position & Credibility level\\
    \hline
        GW190425\_081805 & J124942.30+344928.9 & 0.750 \\
        GW190425\_081805 & J140941.88+552928.1 & 0.555 \\
        GW190425\_081805 & J152433.35+274311.6 & 0.281 \\
        GW190425\_081805 & J163641.61+092459.2 & 0.120 \\
        GW190514\_065416 & J224333.95+760619.2 & 0.479 \\
        GW190519\_153144 & J234420.76+471828.9 & 0.206 \\
        GW190521\_030229 & J124942.30+344928.9 & 0.640 \\
        GW190620\_030421 & J124942.30+344928.9 & 0.747 \\
        GW190620\_030421 & J143536.15+173755.4 & 0.893 \\
        GW190620\_030421 & J163641.61+092459.2 & 0.597 \\
        GW190706\_222641 & J183412.42+365655.2 & 0.685 \\
        GW190708\_232457 & J233252.05+034559.7 & 0.713 \\
        GW190708\_232457 & J233746.08-013116.3 & 0.626 \\
        GW190719\_215514 & J181719.95+541910.0 & 0.727 \\
        GW190731\_140936 & J053408.42+085450.7 & 0.379 \\
        GW190803\_022701 & J053408.42+085450.7 & 0.777 \\
        GW190803\_022701 & J120437.98+500024.0 & 0.544 \\
        GW190403\_051529 & J124942.30+344928.9 & 0.660 \\
        GW190403\_051529 & J140941.88+552928.1 & 0.856 \\
        GW191126\_115259 & J150748.68+723506.1 & 0.888 \\
        GW191129\_134029 & J154342.46+461233.4 & 0.895 \\
        GW191219\_163120 & J154342.46+461233.4 & 0.409 \\
        GW200105\_162426 & J053408.42+085450.7 & 0.790 \\
        GW200105\_162426 & J154342.46+461233.4 & 0.552 \\
        GW200105\_162426 & J154806.31+291216.3 & 0.369 \\
        GW200105\_162426 & J160822.16+401217.8 & 0.819 \\
        GW200112\_155838 & J154342.46+461233.4 & 0.839 \\
        GW200112\_155838 & J160822.16+401217.8 & 0.341 \\
        GW200210\_092255 & J154342.46+461233.4 & 0.836 \\
        GW200216\_220804 & J154342.46+461233.4 & 0.820 \\
        GW200216\_220804 & J154806.31+291216.3 & 0.890 \\
        GW200220\_124850 & J154342.46+461233.4 & 0.145 \\
    \hline
    \end{tabular}
    \label{tab:matches2}
    \end{table}


\section{Method}
\label{sec:method}

The goal of this work is to establish whether the matches between
GW events and AGN flares that are listed in Table \ref{tab:matches1}
and Table \ref{tab:matches2} are due to a causal connection or to
random chance association. To do this, we calculate the posterior
probability distribution of the parameter $f_{\rm flare}$: i.e. of
the fraction of detected mergers that have a causal connection with
a flaring event. As examples of limiting outcomes, if our posterior
would narrowly peak at $f_{\rm flare}=0$ the chance association scenario
would be highly favoured over the physical association scenario, and
viceversa if the narrow peak would be at $f_{\rm flare}=1$.

We use an adapted version of the likelihood presented in
\citet{Veronesi+23}. Such function is based on the
one described in \citet{Braun08}, which has been first re-adapted
to investigate the spatial correlation between GW events and AGN
in \citet{Bartos+17}. The general form is the following:
\begin{align}
    \mathcal{L}&\left(f_{\rm flare}\right)=\prod_{i=1}^{N_{\rm GW}}\mathcal{L}_i\left(f_{\rm flare}\right) \nonumber \\
    &=\prod_{i=1}^{N_{\rm GW}}\left[f_{{\rm cover},i}\cdot0.90\cdot f_{\rm flare}\cdot\mathcal{S}_i
    +\left(1-f_{{\rm cover},i}\cdot0.90\cdot f_{\rm flare}\right)\mathcal{B}_i\right],
    \label{eq:like}
\end{align}
where $N_{\rm GW}$ is the total number of GW detections.
The value of $f_{{\rm cover},i}$ is used to weight the
contribution of the $i$-th single-event likelihood
function. For example, a GW event which sky map is
characterised by $f_{{\rm cover}}=0$ cannot contain
any information regarding the connection with AGN flares
observed by ZTF. Using $f_{{\rm cover},i}$ as weight in the
likelihood calculation takes this into account. Analyses
that investigate the spatial correlation between GW events and
non-transient potential EM counterparts use as a weight for the
single-event likelihoods the completeness of the AGN catalogue
instead of $f_{{\rm cover},i}$ \citep[see, for example][]{Veronesi+23}.
The role of these parameters in these similar analyses is the same.

The signal probability density for the $i$-th GW
event is calculated as follows:
\begin{equation}
    \mathcal{S}_i=\sum_{j=1}^{N_{{\rm matches}_i}}\frac{p_j}{n_{{\rm flare},j}}
    \frac{1}{{\rm V90}_i},
    \label{eq:Si}
\end{equation}
where ${N_{{\rm matches}_i}}$ is the number of spatial
and temporal matches the $i$-th GW event has with flares,
$p_j$ is the probability density associated to the
position of the $j$-th matching EM transient, and $n_{{\rm flare},j}$
is the effective spatial number density correspondent
to it. This number density is assumed to be uniform within V90.

In the analysis that involves the 2D sky-projected cross-match,
${\rm A90}$ is used instead of ${\rm V90}$, and both $p_j$ and
$n_{{\rm flare},j}$ are in units of ${\rm deg}^{-2}$, and
not in units of ${\rm Mpc}^{-3}$. We calculate the effective
number density as follows:
\begin{equation}
    n_{{\rm flare},j}=\frac{20}{V_{\rm ZTF, eff}}\frac{200-t_{{\rm g},j}}{\Delta t_{\rm search}},
    \label{eq:nflare}
\end{equation}
where 20 is the total number of AGN flares selected from
ZTF DR5 as potential counterparts of GW events,
$V_{\rm ZTF, eff}=1.066 \times 10^{11}{\rm Mpc}^3$ is the
total effective surveyed comoving volume, and
$\Delta t_{\rm search}=562\ {\rm days}$ is the total
observing time of the search, between the start of O3 and
200 days after its end. We calculate the total effective
comoving volume in order for it to be compatible with
how $f_{\rm cover}$ has been calculated for each GW
event, and to take into account the non uniformity of
the source density in the ZTF footprint. We first map
the sky into a HealPix grid with a resolution of NSIDE=32.
To each pixel correspondent to a galactic latitude
lower than $-10^\circ$ or greater than $10^\circ$ is
then associated a number between 0 and 1. This
value is the fraction of the total number of GW events
for which that specific pixel was observed at least
20 times in both the ZTF g-band and r-band in the
200 days following the merger detection. These values
are then all summed together and the result is multiplied
by the angular size of each pixel. This sum corresponds
to the total effective area observed by the survey during
$\Delta t_{\rm search}$, $A_{\rm ZTF, eff}$, and is
used in the calculation of $n_{{\rm flare},j}$, instead
of $V_{\rm ZTF, eff}$, in the case of the 2D sky-projected
analysis. The value of $V_{\rm ZTF, eff}$ is finally
calculated multiplying the total comoving volume enclosed
within the redshift limit adopted in the search presented
in G23, $z=1.2$, by the ratio between $A_{\rm ZTF, eff}$
and the total area of the sky. Finally, the numerator
of the second fraction in Equation \ref{eq:nflare} takes
into account the width of the time window within which
a specific flare has to peak in order to allow the
match with a GW event. We calculate this quantity by
subtracting the Gaussian rise time of the flare from
200 days, the maximum allowed delay time between the
peak of such EM transient and the GW detection.

The background probability density function $\mathcal{B}_i$
is calculated in the same way as in \citet{Veronesi+23}:
\begin{equation}
    \mathcal{B}_i=\frac{0.9}{{\rm V90}_i} ,
    \label{eq:Bi}
\end{equation}
where the 0.9 factor is used so that $\mathcal{S}_i$ and
$\mathcal{B}_i$ are normalised to the same value. Just
like the signal density function (Equation \ref{eq:Si})
${\rm A90}$ is used instead of ${\rm V90}$ for the 2D
cross-matching analysis.

Every single-event likelihood $\mathcal{L}_i\left(f_{\rm flare}\right)$
is a monotonic function of $f_{\rm flare}$, the only free parameter.
For each merger, if $\mathcal{S}_i>\mathcal{B}_i$, then
$\mathcal{L}_i\left(f_{\rm flare}=1\right)>\mathcal{L}_i\left(f_{\rm flare}=0\right)$;
this means that for that specific GW event our spatial correlation
analysis favours the hypothesis according to which there is a
causal connection with AGN flares. This can happen if there is an
over-density of temporally-matching flaring activities inside
the localisation volume of the merger, and/or if the potential
counterparts are located in regions of the sky map with a high
value of the probability density $p_j$. In general, the greater
the difference between $\mathcal{L}_i\left(f_{\rm flare}=1\right)$
and $\mathcal{L}_i\left(f_{\rm flare}=0\right)$, the more information
the single-event likelihood brings to the total one.

The main distinction between the analysis presented in this
work and previous ones, that involve cross-matching GW sky maps with
regular AGN, is that flares are transient events. This is taken into
account in different parts of our statistical framework. First and
foremost, a temporal match
is required between GW detections and observed AGN flaring activities
in order for them to be considered as potentially related by causality.
Moreover, each value of $n_{\rm flare}$ depends on the width of the temporal
window that allows for a match, and on $V_{\rm ZTF, eff}$. The latter takes
into account the scanning pattern of ZTF and how much it covers the
localisation areas of the mergers in the days after the detection.
Finally, each value of $f_{\rm cover}$ accounts for transient nature of
the potential EM counterparts considered in this work (see Section
\ref{subsec:gws}).

After normalising $\mathcal{L}\left({\rm f_{\rm flare}}\right)$
using a flat prior on $f_{\rm flare}$ in the $[0,1]$ interval,
we obtain the posterior distribution of this parameter.


\section{Results}
\label{sec:res}

In this Section we present the posterior distribution of
the fraction of detected GW events that have a connection
with an AGN flare, $f_{\rm flare}$. Then we show the
results of two more tests to check if this posterior is
indeed compatible with the $f_{\rm flare}=0$ hypothesis.
The first  of the two tests focuses the total number of
matches between the catalogue of AGN flares and the one
of GW events, the second on the binary mass distribution
one should expect for the mergers characterised by at
least one match.

\subsection{Posterior distributions}
\label{subsec:posts}
\begin{figure}
    \centering
    \includegraphics[trim= 0 0 0 0 ,clip,width=0.49\textwidth]{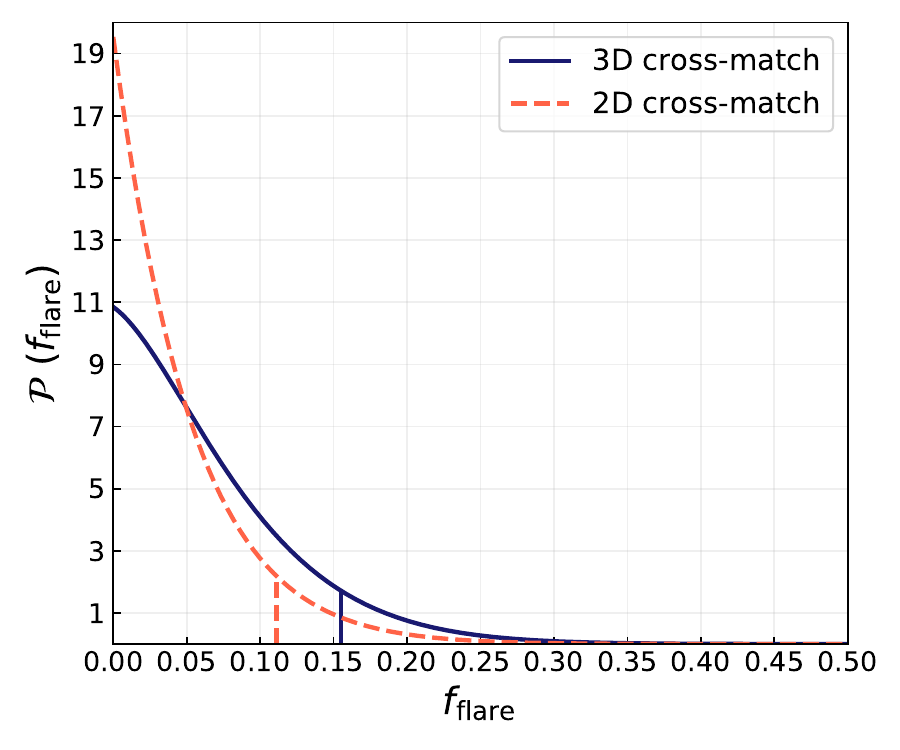}
    \caption{Posterior probability distribution on the fraction
    of GW events that are expected to have caused a flare in an
    AGN, $f_{\rm flare}$. The solid blue line corresponds to the
    result of the 3D cross-matching analysis, while the dashed
    pink line to the one of the 2D sky-projected case. Both
    functions peak at ${{f}_{\rm flare}=0}$. The vertical lines
    indicate the upper limits of the 90 per cent Credibility
    Intervals. These correspond to ${f_{\rm flare}=0.155}$ and
    ${f_{\rm flare}=0.111}$ for the 3D and 2D cross-matching
    analysis, respectively.}
    \label{fig:post}
\end{figure}

The posterior probability distribution on $f_{\rm flare}$ is
estimated using the data presented in Section \ref{sec:dataset}
and the method described in Section \ref{sec:method}. The result
is shown in Figure \ref{fig:post}. For both the 3D and the
2D cross-matching analyses, the posterior distribution peaks
at ${{f}_{\rm flare}=0}$. In the case of the 3D analysis,
the upper limits of the 68, 90, and 95 per cent Credibility
Intervals (CIs) correspond to ${f_{\rm flare}=0.085}$,
${f_{\rm flare}=0.155}$, and ${f_{\rm flare}=0.192}$, respectively.
The upper limits for the same CIs in the 2D sky-projected analysis are
${f_{\rm flare}=0.056}$, ${f_{\rm flare}=0.111}$, and ${f_{\rm flare}=0.142}$.

We estimate wether or not a model according to which there is no
relation between the detected GW events and the observed flares is
preferred with respect to a model according to which $f_{\rm flare}$ can
assume any value in the $\left[0,1\right]$ range. To do so we calculate the
Bayes factor defined as follows:
\begin{align}
    \mathcal{K}&=\frac{\int_0^1\mathcal{L}\left(f_{\rm flare}\right)\delta\left(f_{\rm flare}-0\right){\rm d}f_{\rm flare}}{\int_0^1\mathcal{L}\left(f_{\rm flare}\right)\pi\left(f_{\rm flare}\right){\rm d}f_{\rm flare}} \nonumber \\
    &=\frac{\mathcal{L}\left(f_{\rm flare}=0\right)}{\int_0^1\mathcal{L}\left(f_{\rm flare}\right)\pi\left(f_{\rm flare}\right){\rm d}f_{\rm flare}}\ \ ,
\end{align}
where $\delta\left(f_{\rm flare}-0\right)$ is a Dirac delta distribution,
and the prior function $\pi\left(f_{\rm flare}\right)$ is uniform in the
$\left[0,1\right]$ range. In the case of the 3D and the 2D analysis,
$\mathcal{K}\approx11.59$ and $\mathcal{K}\approx19.55$, respectively.
These values show a strong preference for the model according to which
$f_{\rm flare}=0$ \citep{Jeffreys61,Kass_Raftery95}.


\subsection{Background Monte Carlo realisations}
The posterior distributions presented in Section \ref{subsec:posts}
peak at ${{f}_{\rm flare}=0}$, which corresponds to the
hypothesis of no causal connection between mergers detected
by the LVK interferometers and the AGN flares selected as potential
EM counterparts for such events. In other words, the positioning
of the flares within the GW sky maps, and the observed
number of matches between GW events and AGN flares appear
consistent with chance associations only, without any physical
relation between the two events.

\begin{figure}
    \centering
    \includegraphics[trim= 10 0 0 32 ,clip,width=0.52\textwidth]{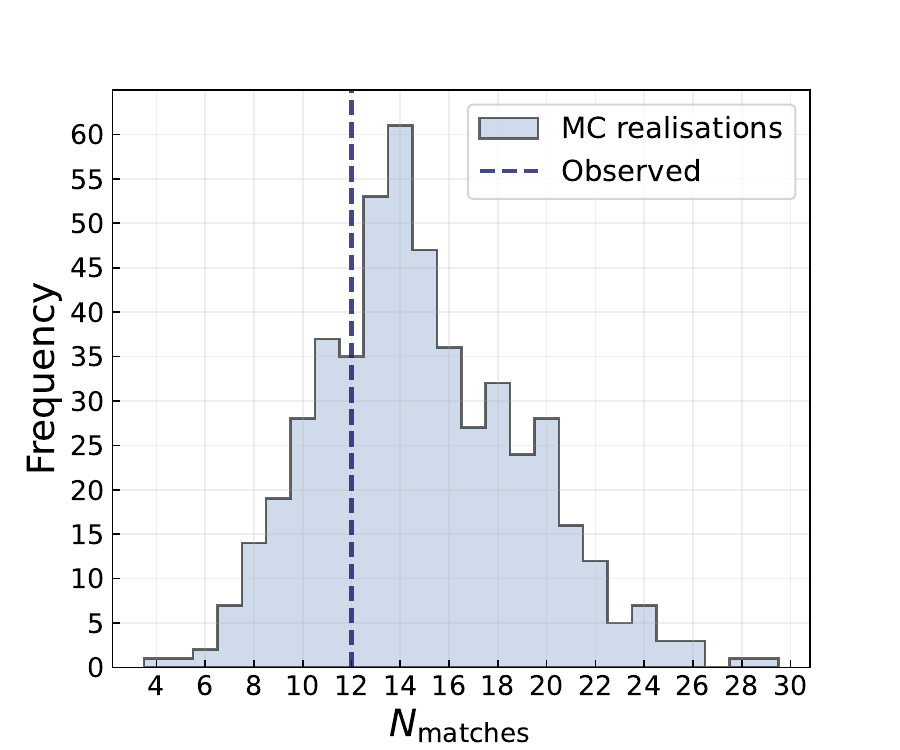}
    \caption{Distribution of the number of spatial and temporal
    matches between simulated ZTF AGN flares and GW 3D sky maps
    obtained from 500 MC realisations. Each of these realisations
    represents the scenario in which there is no causal connection
    between the two different signals, and every match is due to
    random chance association. The average of the distribution
    is $\bar{N}_{\rm matches}=14.9$, and its standard deviation is
    $\sigma\left(N_{\rm matches}\right)=4.2$. The vertical dashed
    line indicates the number of matches that have been found
    using real observed data (see Table \ref{tab:matches1}).}
    \label{fig:NMC}
\end{figure}

To further test this hypothesis, we perform 500 Monte Carlo
(MC) realisations of this background scenario. Each of them is
constructed as follows:
\begin{itemize}
    \item To generate a catalog of AGN flares that follows the ZTF
    real sampling of the sky, which is not fully uniform within
    the footprint, we sample 20 sky-positions from the
    catalogue of extra-galactic transient events presented in
    \citet{vanVelzen24};
    \item For each simulated flare, we select its peak time by
    drawing from the times its sky position was observed by ZTF
    To match the peak time distribution of the 20 flares from G23, we
    require this peak time to be later than the start of O3 and
    earlier than the date that corresponds to 200 days after
    the end of the same observing run;
    \item To each of the 20 sampled positions we associate
    a value of redshift. This is obtained by inverse-sampling a
    linear interpolation of the Cumulative Distribution Function
    of the redshifts of the potential EM counterparts to GW
    events listed in Table \ref{tab:flares};
    \item To each flare we associate a value for its rise
    time ${t_{\rm g}}$, drawn randomly from the ones listed in
    Table \ref{tab:flares};
    \item Once the catalogue of simulated background ZTF flares is
    constructed, it is cross-matched with the 3D sky maps of the
    GW events detected during O3 and listed in Table \ref{tab:gws}.
    As for the cross-match performed with observed data, in order
    for a match to be considered as valid, the flare has to peak
    not more than 200 days after the GW detection, and not before
    that a number of days equal to the corresponding ${t_{\rm g}}$
    have passed.
\end{itemize}
The histogram in Figure \ref{fig:NMC} shows the distribution of the
number of spatial and temporal matches obtained in the 500 MC
realisations of the background scenario. The average value of the
sample is $\bar{N}_{\rm matches}=14.9$ and its standard deviation
is $\sigma\left(N_{\rm matches}\right)=4.2$. The number of matches
obtained from observed data (12) is less than one standard deviation
away from the mean of the distribution. It is reasonable then to
assess that 12 matches can be expected even only due to random chance
association.

For each MC realisation, the posterior distribution of
$f_{\rm flare}$ is also obtained. In 150 realisations this
function peaks at ${f}_{\rm flare}=0$, and the distribution
of the value of ${f_{\rm flare}}$ extends to
${f}_{\rm flare}\approx0.26$. This is similar to the posterior
of $f_{\rm fare}$ obtained from observed data, again confirming
that the outcome of our likelihood method (Eq.\ref{eq:like})
is consistent with the background hypothesis.


\subsection{Binary Mass distributions}
\begin{figure*}
    \centering
    \includegraphics[trim= 0 0 0 0 ,clip,width=0.75\textwidth]{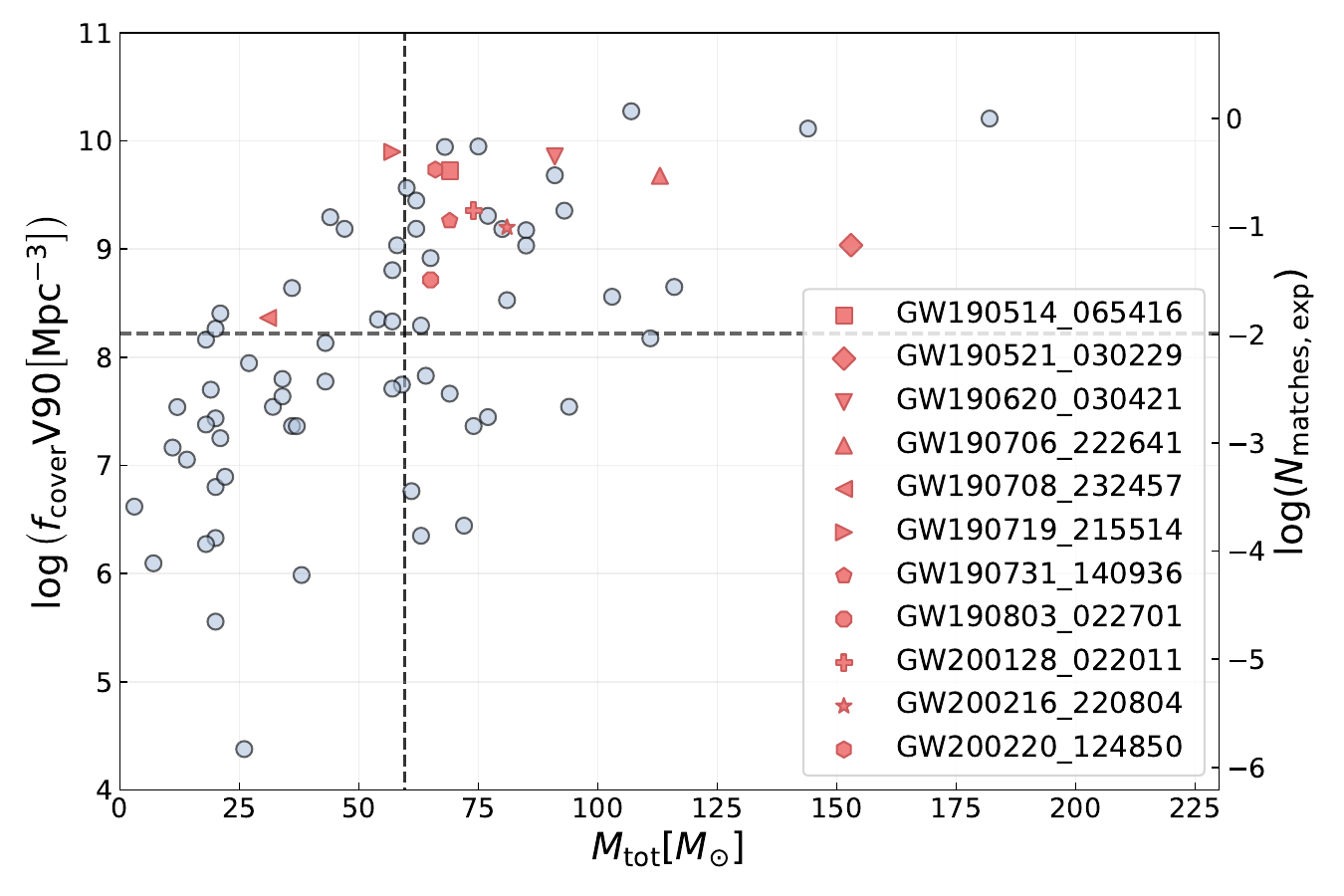}
    \caption{Effective size of V90 of the GW events detected
    during O3 as a function of the source-frame total mass of
    the binary. The 11 pink non-round markers indicate the
    mergers that have a spatial and temporal match with at least
    one of the AGN flares listed in Table \ref{tab:flares}.
    These matching sources appear to have a large mass compared
    to the median of the entire population (indicated with the
    dashed lines). However this can be explained by the correlation
    between mass and volume, see also Figure \ref{fig:ksmass}.
    On the right-hand side y-axis is indicated the expectation
    number of random matches evaluated assuming a uniform average
    effective number density calculated from Equation \ref{eq:nflare},
    using the mean $t_{\rm g}$ of the AGN flares.}
    \label{fig:v90vsmass}
\end{figure*}

\begin{figure}
    \centering
    \includegraphics[trim= 5 0 0 0 ,clip,width=0.49\textwidth]{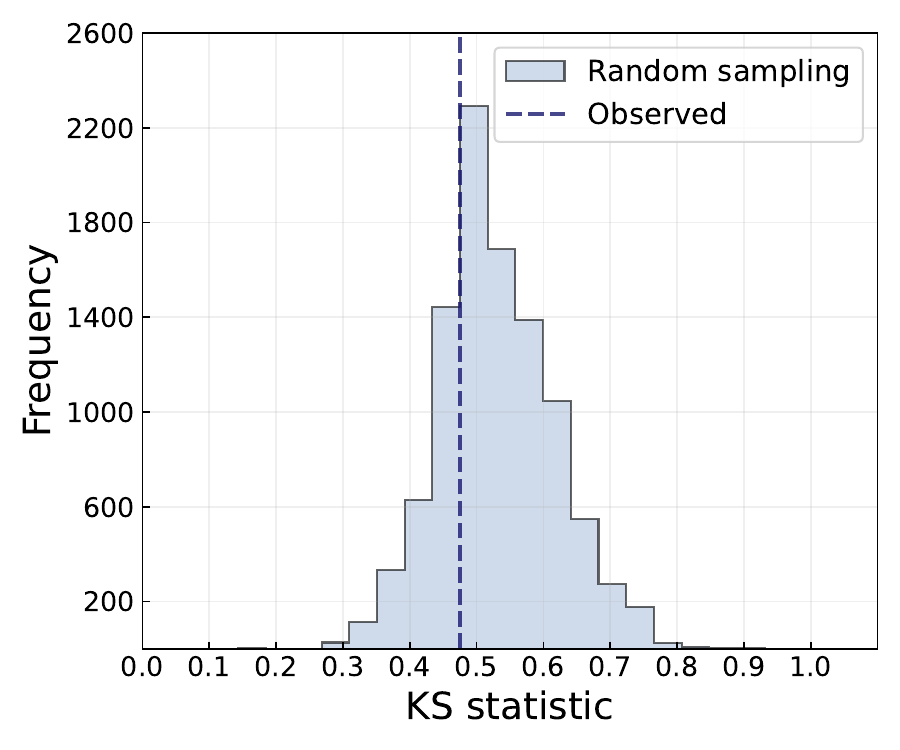}
    \caption{Distribution of the KS statistic obtained by
    comparing the source-frame total binary masses of GW events
    that have a 3D spatial match with an AGN flare with the ones
    of the other mergers detected during O3. The histogram shows
    the distribution of such statistic obtained from 10,000
    repetitions of random sampling, where the number of matches
    of every GW event was drawn from a Poisson distribution
    that depends on the corresponding size of V90. The vertical
    dashed blue line indicates the value of the KS statistic
    calculated with the observed data. After correcting for
    the correlation between mass and V90, we thus find no
    evidence for a difference between the mass distribution
    of the GW sources that match to AGN flares and the mass
    distributions of the ones that do not.}
    \label{fig:ksmass}
\end{figure}

Due primarily to mass segregation and the possibility of
undergoing subsequent hierarchical mergers, BBHs that coalesce
inside dynamically dense environments like the accretion
discs of an AGN are expected to have on average a higher
mass with respect to the ones that formed through the
evolution of an isolated binary stellar system. In Figure
\ref{fig:v90vsmass} it is shown the size of V90 for each
GW event detected during O3, weighted by $f_{\rm cover}$,
as a function of the source-frame total mass of the merging
binary. The horizontal and the vertical dashed lines indicate
the median weighted size of V90 and the median total mass,
respectively. The pink non-round markers indicate the 11 GW
events that have at least one 3D match with a potential EM
counterpart. 
We indeed notice that most of matching GW events have a
total mass that is higher than the median. A KS test for
the hypothesis that the mass distributions of the matching
and the non-matching merging binaries are not related
yields $p=0.018$. 
However, this apparent difference in mass might be because
the matching events are also associated with localisation
uncertainties larger than the median.
To test this hypothesis we follow the procedure presented hereunder:
\begin{itemize}
    \item The average effective flare number density,
    $\langle n_{\rm flare}\rangle$, is first calculated
    from Equation \ref{eq:nflare}, using the average rise time of
    the flares listed in Table \ref{tab:flares},
    $\langle t_{\rm g}\rangle=17.35$;
    \item For all the GW events we evaluate the expected number of
    matches due to chance association, $N_{\rm matches, exp}$,
    multiplying the size of their localisation volume by
    $\langle n_{\rm flare}\rangle$;
    \item We extract a number of matches for each GW event, drawing
    from a Poisson distribution with $N_{\rm matches, exp}$ as
    expectation value;
    \item We finally perform a two-sample Kolmogorov-Smirnov test
    between the distribution of the source-frame total mass of
    the GW events that have at least one match in this random
    sample and the distribution of the source-frame mass of the
    ones that don't have any match. For each repetition we therefore
    obtain a value of the KS statistic.
\end{itemize}
The whole process is repeated 10,000 times. The histogram in Figure
\ref{fig:ksmass} shows the distribution of the KS statistic for all
the repetitions of the random sampling. We see that the KS statistic
of the observed 11 matching GW events is consistent with the distribution
obtained from the random sampling process. This means that the fact
that the observed matching events correspond on average to higher
binary masses with respect to the non-matching ones can be
explained assuming random chance associations between GW sky maps
and the AGN flares, if the positive correlation that exists between
source-frame binary total masses and the size of the reconstruction
volumes is taken into account.

\section{Discussion and conclusion}
\label{sec:concl}
We present a statistical investigation on the connection between the
GW events detected during O3 and the AGN flares selected in G23 as
potential EM counterparts. We do this using the most updated version
of the posterior samples released by the LVK collaboration. We make
use of the statistical method presented in \citet{Veronesi+23} to
estimate the posterior distribution over the fraction of GW events
that have an observed AGN flare as an EM counterpart, $f_{\rm flare}$.
We repeat the same analysis both performing a 3D cross-match between
the GW sky maps and the positions of the flares, and a 2D sky-projected
one. In both cases the posterior distribution peaks at
${f}_{\rm flare}=0$.

The upper limit of the 90 per cent CI is $f_{\rm flare}=0.155$ in the
case of the 3D cross-matching analysis and $f_{\rm flare}=0.111$ for
the 2D sky-projected case.

We calculate the Bayes factor to compare a model in which $f_{\rm flare}=0$
to a model in which such parameter can assume any value in the $\left[0,1\right]$
range. We find the former to be strongly preferred with respect to the
latter both in the case of the 3D and the 2D analyses.

Moreover, we perform 500 MC realisations of the background hypothesis,
according to which the matches between GW events and AGN flares are
due to random chance association. We find that the total number of
matches that exist in the case of the real observed data (12) is
compatible with this scenario.

Finally we find through a random sampling process, that the observed
distribution of source-frame binary masses of the mergers that have
a spatial and temporal match with at least one AGN flare is compatible
with the same no-connection hypothesis.

We conclude that the hypothesis of no causal connection between the
detected GW events and the observed AGN flares cannot be rejected.

In G23 Poisson statistic is used to calculate the probability of
obtaining a number of matches equal or bigger than the observed one
under the no causal connection hypothesis. They estimate this
probability to be $p=0.0019$. The estimate for the number of expected
random matches used in G23 (2.83) is calculated assuming
a spatial number density uniform in comoving volume up to $z=1.2$
and a temporal distribution of the flares over the whole duration
of the data collection of ZTF DR5.
This estimate is not compatible with the average number of matches
we obtain in our Monte Carlo realisations of the background hypothesis
(14.9). A key difference is that we take into account non-uniform
distributions of the sky-position (by sampling from ZTF sources to
obtain a realistic sky distributions) and redshift of the AGN flares
(by sampling from the redshifts of the observed flares), and we allow
these transient events to peak only in the time window that includes
all the 20 selected potential counterparts. Below we explain this
difference in more detail. 

The main distinction between this work and G23 is that we measure a
higher effective AGN flare number density, which leads to a higher expected
number of matches under the background hypothesis. This difference is
mainly driven by two factors. First, in G23 the fraction of matching
temporal windows is calculated assuming a mean flare lifetime of 100 days
and dividing it by 1000 days, approximately the duration of the
observations of ZTF DR5.
However, all the 20 AGN flares considered in the analysis peaked between
the start of O3 and 200 days after its end, implying the effective duration of
the ZTF flare search is 562 days. As such, we calculate for each flare
the fraction of matching temporal windows by dividing the width of
this window ($200\ {\rm days} - t_{\rm g}$) by 562 days. This leads
to a significant difference with respect to the value assumed in G23,
since the average value of the rise times, $t_{\rm g}$, is about 17
days. The second factor is a different estimate of the effective total
comoving volume probed in the ZTF search. Our estimate
($V_{\rm ZTF, eff}=1.066 \times 10^{11}{\rm Mpc}^3$) takes into account
the non-uniformity of the sky distribution of the ZTF extra-galactic
sources and is approximately 1.5 times smaller compared to the ZTF
volume used in G23. Taken together these two effects yield an effective
source density that is approximately 4.6 times larger than the one used in G23.
We find a value of the total localisation volume weighted by $f_{\rm cover}$
similar to the ones used in G23, even using updated sky maps
($1.457\cdot10^{11}{\rm Mpc}^3$, that corresponds approximately to 68
per cent of the value used in G23).

In \citet{Palmese21} it is found that there is a 70 per cent
probability of chance coincidence for a flaring event to happen within
the 90 per cent localisation volume of GW190521\_030229. This result
is obtained using their estimate for the number density of AGN flares
with a $g$ magnitude variation of $\Delta m\geq 0.4$ and the most
up-to-date estimate of V90 for GW190521\_030229 available when
the analysis was performed. They conclude that the hypothesis
according to which the spatial and temporal match between the GW event
and the flaring activity in the AGN J124942.3+344929 is not the result
of a causal connection between the two detected transients cannot be
confidently rejected. They reach the same conclusion even when using
a different estimate of the AGN flares number density. This second estimate
is derived from the chance the flare model presented in \citet{Graham20}
has to fit any ZTF AGN light-curve. In this case the probability of
chance association is found not to go below $\approx4$ per cent.
By multiplying the same value of V90 used in \citet{Palmese21} for
GW190521\_030229 by the average number density of unusual AGN flares
obtained by using $\langle t_{\rm g}\rangle=17.35$ in Equation
\ref{eq:nflare} of this work, we find a probability of $\approx 25$ per
cent of chance
association between the GW event and an EM transient like the ones
listed in Table \ref{tab:flares}. This result, obtained using the
estimate of the number density of AGN flares that enters our
analysis, is in agreement with the conclusion stated in \citet{Palmese21}
regarding the impossibility to confidently associate
GW190521\_030229 with the flare of J124942.3+344929.

It is evident that our analysis only constraints the fraction of GW
events in AGN that yield a detectable flare. Several factor can make
this counterpart challenging to detect.

In order to produce a potentially
observable flare not later than 200 days after the GW event, the
recoil velocity has to be greater than a value that depends on the
physical and geometrical characteristics of the AGN disc (see Equation
5 of G23). At the same time, an increase in the recoil velocity corresponds
to a decrease in the Bondi-Hoyle-Lyttleton luminosity (see Equations
3 and 6 of G23). Therefore if the merger remnant travels
too fast through the accretion disc, the hypothetical resulting flare
will not be detected as it will not significantly exceed the luminosity
of the host AGN. The recoil velocity of the merger
remnant plays a role also in the jetted model of a BBH-induced AGN flare
described in \citet{Tagawa+22} and in \citet{Tagawa+23b}. In particular,
the rate at which the BH captures the gas, and therefore the luminosity
of the jet, are expected to increase as a function of the velocity of
such compact object with respect to the local motion of the disc \citep[see
Equation 1 of][]{Tagawa+22}. However, the accretion rate is expected to
be reduced when the radius at which there can be gas accretion onto the
BH on the timescale of the breakout of the jet at the surface of the AGN
disc is much bigger than the radius at which the gaseous material remains
bound to the compact object after the recoil kick. This is due to the
ejection of gas beyond the bound region surrounding the merger remnant,
and can happen if the recoil velocity is high ($\gtrsim2000\ {\rm km\ s}^{-1}$
in the case of a BH with a mass of $150{\rm M}_\odot$) \citep[see
Appendix B of] [for a detailed description of the process]{Tagawa+23b}.

On top of these physical factors regarding the observability of the
flare caused by a kicked remnant
exiting the accretion disc, there is a geometrical one. To be
visible from instruments like ZTF, the flare has in fact to happen
on the side of the disc that faces Earth, and on average this is
expected to be true only in half of the cases.

Finally, in order to be confidently identified as a potential EM
counterpart to a GW signal, the flare has to happen in a position
in the sky that allows the light-curve to be scanned with several
observations in different bands over its lifetime.

Current and future time-domain surveys such as ZTF and
the Vera C. Rubin Observatory \citep{Ivezic19} are promising tools
for the identification of transient EM counterparts of GW events
detected by the LVK collaboration. The low number density of
transient EM events not compatible with regular AGN variability
makes them ideal for spatial correlation analyses like the one
presented in this work and for live searches of counterpart
candidates of specific GW events, like the one presented in
\citet{Cabrera24}, which focuses on the BBH merger candidate
S230992g detected during the fourth observing run of the LVK
collaboration.

In the population analysis presented in this work every
AGN flare is considered as a potential counterpart of all the
GW events with which it has a spatial and temporal match. This
leads to some of these EM transients to be considered in our calculations more than
once even if, from a physical point of view, each of them can be causally
related to at most on BBH merger. Future different versions of
spatial correlation analyses can be developed and applied to
observed data, assigning every flare to at most one GW event, such
as the one the sky map of which has the highest value of probability
density in the position of the flaring AGN.

Current data suggest a lack of confident causal connection
between observed unusual AGN flares and the
events detected by the LVK interferometers, but updating
the posterior distribution of $f_{\rm flare}$ with more data,
or data more sensitive to lower amplitude flares, could
yield better constraints on the relation between mergers
of compact objects and AGN flares. 


\vspace{-0.8em}
\section*{Acknowledgements}
The authors thank Matthew Graham for making the data of the 20 AGN
flares mentioned in G23 available, and the anonymous referees,
whose comments have helped to ameliorate the clarity of the
presentation of our results.
EMR acknowledges support from ERC Grant ``VEGA P.", number 101002511.
This research has made use of data or software obtained from
the Gravitational Wave Open Science Center (gwosc.org), a service
of LIGO Laboratory, the LIGO Scientific Collaboration, the Virgo
Collaboration, and KAGRA. LIGO Laboratory and Advanced LIGO are
funded by the United States National Science Foundation (NSF) as
well as the Science and Technology Facilities Council (STFC) of
the United Kingdom, the Max-Planck-Society (MPS), and the State
of Niedersachsen/Germany for support of the construction of Advanced
LIGO and construction and operation of the GEO600 detector.
Additional support for Advanced LIGO was provided by the Australian
Research Council. Virgo is funded, through the European Gravitational
Observatory (EGO), by the French Centre National de Recherche Scientifique
(CNRS), the Italian Istituto Nazionale di Fisica Nucleare (INFN)
and the Dutch Nikhef, with contributions by institutions from Belgium,
Germany, Greece, Hungary, Ireland, Japan, Monaco, Poland, Portugal,
Spain. KAGRA is supported by Ministry of Education, Culture, Sports,
Science and Technology (MEXT), Japan Society for the Promotion of
Science (JSPS) in Japan; National Research Foundation (NRF) and
Ministry of Science and ICT (MSIT) in Korea; Academia Sinica (AS) and
National Science and Technology Council (NSTC) in Taiwan.
{\em Software}: 
\texttt{Numpy} \citep{Harris20}; 
\texttt{Matplotlib} \citep{Hunter07}; 
\texttt{SciPy} \citep{Virtanen20};
\texttt{Astropy} \citep{Astropy13,Astropy18};
\texttt{BAYESTAR} \citep{Singer16}.

\vspace{-0.8em}
\section*{Data Availabilty}
The data underlying this article will be shared on reasonable
request to the corresponding author.

\vspace{-1.0em}
\bibliographystyle{mnras}
\bibliography{bibliography}

\appendix
\section{List of GW events}
\label{app:gws}
In Table \ref{tab:gws} we list the properties of the GW events
detected during O3 we used in the analysis presented in this work.
\vfill
\onecolumn
\begin{longtable}[c]{ccrccc}
\caption{List of GW events used in the analysis presented in this
work. For each of them we list the ID, the catalogue it is contained
in, the size of its 90 per cent credibility localisation area and
volume, the Modified Julian Day of the detection, and the fraction of
its 90 per cent credibility level localisation area that has been
observed by ZTF at least 20 times in both the g-band and r-band during
200 days following the GW detection, at a galactic latitude $|b|>10^\circ$.}

\label{tab:gws}\\
    \hline
    GW ID & Catalogue & A90 & V90 & MJD & $f_{\rm cover}$\\
     &  &$\left[{\rm deg}^{2}\right]$ & $\left[{\rm Mpc}^3\right]$ & & \\
\hline
\endfirsthead
    \hline
    GW ID & catalogue & A90 & V90 & MJD & $f_{\rm cover}$\\
     & & $\left[{\rm deg}^{2}\right]$ & $\left[{\rm Mpc}^3\right]$ & & \\
\hline
\endhead
\hline
\endfoot
         GW190403\_051519 & GWTC-2.1 & 2731 & 3.872$\cdot10^{10}$ & 58576.2 & $0.487$\\
         GW190408\_181802 & GWTC-2.1 & 271 & 8.922$\cdot10^7$ & 58581.8 & $0.670$\\
         GW190412\_053044 & GWTC-2.1 & 25 & 1.112$\cdot10^6$ & 58585.2 & $0.875$\\ 
         GW190413\_052954 & GWTC-2.1 & 668 & 2.017$\cdot10^9$ & 58586.2 & $0.539$\\
         GW190413\_134308 & GWTC-2.1 & 562 & 2.100$\cdot10^9$ & 58586.6 & $0.161$\\
         GW190421\_213856 & GWTC-2.1 & 1237 & 1.729$\cdot10^9$ & 58594.9 & $0.013$\\
         GW190425\_081805 & GWTC-2.1 & 8728 & 7.772$\cdot10^6$ & 58598.3 & $0.537$\\
         GW190426\_190642 & GWTC-2.1 & 4559 & 2.527$\cdot10^{10}$ & 58599.8 & $0.638$\\
         GW190503\_185404 & GWTC-2.1 & 103 & 4.716$\cdot10^7$ & 58606.8 & $0.000$\\
         GW190512\_180714 & GWTC-2.1 & 274 & 9.283$\cdot10^7$ & 58615.8 & $0.250$\\
         GW190513\_205428 & GWTC-2.1 & 448 & 3.805$\cdot10^8$ & 58616.9 & $0.590$\\
         GW190514\_065416 & GWTC-2.1 & 3186 & 1.063$\cdot10^{10}$ & 58617.3 & $0.515$\\
         GW190517\_055101 & GWTC-2.1 & 365 & 3.042$\cdot10^8$ & 58620.2 & $0.222$\\
         GW190519\_153544 & GWTC-2.1 & 672 & 1.180$\cdot10^9$ & 58622.6 & $0.308$\\
         GW190521\_030229 & GWTC-2.1 & 1021 & 3.434$\cdot10^9$ & 58624.1 & $0.500$\\
         GW190521\_074359 & GWTC-2.1 & 469 & 5.611$\cdot10^7$ & 58624.3 & $0.500$\\
         GW190527\_092055 & GWTC-2.1 & 3640 & 7.788$\cdot10^9$ & 58630.4 & $0.473$\\
         GW190602\_175927 & GWTC-2.1 & 739 & 1.412$\cdot10^9$ & 58636.7 & $0.317$\\
         GW190620\_030421 & GWTC-2.1 & 6443 & 1.219$\cdot10^{10}$ & 58654.1 & $0.618$\\
         GW190630\_185205 & GWTC-2.1 & 960 & 1.104$\cdot10^8$ & 58664.8 & $0.507$\\
         GW190701\_203306 & GWTC-2.1 & 43 & 3.494$\cdot10^7$ & 58665.9 & $1.000$\\
         GW190706\_222641 & GWTC-2.1 & 2596 & 7.799$\cdot10^9$ & 58670.9 & $0.590$\\
         GW190707\_093326 & GWTC-2.1 & 893 & 6.443$\cdot10^7$ & 58671.4 & $0.098$\\
         GW190708\_232457 & GWTC-2.1 & 11032 & 9.846$\cdot10^8$ & 58673.0 & $0.533$\\
         GW190719\_215514 & GWTC-2.1 & 3564 & 1.317$\cdot10^{10}$ & 58683.9 & $0.612$\\
         GW190720\_000836 & GWTC-2.1 & 35 & 2.303$\cdot10^6$ & 58684.0 & $0.000$\\
         GW190725\_174728 & GWTC-2.1 & 2142 & 3.780$\cdot10^8$ & 58689.7 & $0.386$\\
         GW190727\_060333 & GWTC-2.1 & 100 & 1.963$\cdot10^8$ & 58691.3 & $0.235$\\
         GW190728\_064510 & GWTC-2.1 & 321 & 2.974$\cdot10^7$ & 58692.3 & $0.602$\\
         GW190731\_140936 & GWTC-2.1 & 3532 & 8.919$\cdot10^9$ & 58695.6 & $0.357$\\
         GW190803\_022701 & GWTC-2.1 & 1012 & 2.227$\cdot10^9$ & 58698.1 & $0.684$\\
         GW190805\_211137 & GWTC-2.1 & 1538 & 1.342$\cdot10^{10}$ & 58700.9 & $0.664$\\
         GW190814\_211039 & GWTC-2.1 & 22 & 3.590$\cdot10^4$ & 58709.9 & $0.667$\\
         GW190828\_063405 & GWTC-2.1 & 340 & 2.502$\cdot10^8$ & 58723.3 & $0.205$\\
         GW190828\_065509 & GWTC-2.1 & 593 & 2.700$\cdot10^8$ & 58723.3 & $0.234$\\
         GW190910\_112807 & GWTC-2.1 & 8305 & 5.158$\cdot10^9$ & 58736.5 & $0.394$\\
         GW190915\_235702 & GWTC-2.1 & 432 & 2.417$\cdot10^8$ & 58742.0 & $0.889$\\
         GW190916\_200658 & GWTC-2.1 & 2368 & 1.537$\cdot10^{10}$ & 58742.8 & $0.574$\\
         GW190917\_114630 & GWTC-2.1 & 1687 & 1.096$\cdot10^8$ & 58743.5 & $0.317$\\
         GW190924\_021846 & GWTC-2.1 & 376 & 1.209$\cdot10^7$ & 58750.1 & $0.939$\\
         GW190925\_232845 & GWTC-2.1 & 876 & 9.957$\cdot10^7$ & 58752.0 & $0.232$\\
         GW190926\_050336 & GWTC-2.1 & 2015 & 7.945$\cdot10^9$ & 58752.2 & $0.355$\\    
         GW190929\_012149 & GWTC-2.1 & 1651 & 4.851$\cdot10^9$ & 58755.1 & $0.468$\\
         GW190930\_133541 & GWTC-2.1 & 1493 & 1.223$\cdot10^8$ & 58756.6 & $0.223$\\
         GW191103\_012549 & GWTC-3 & 2171 & 2.663$\cdot10^8$ & 58790.1 & $0.692$\\
         GW191105\_143521 & GWTC-3 & 641 & 1.250$\cdot10^8$ & 58792.6 & $0.402$\\
         GW191109\_010717 & GWTC-3 & 1649 & 4.863$\cdot10^8$ & 58796.0 & $0.308$\\
         GW191113\_071753 & GWTC-3 & 2484 & 1.159$\cdot10^9$ & 58800.3 & $0.378$\\
         GW191126\_115259 & GWTC-3 & 1378 & 5.990$\cdot10^8$ & 58813.5 & $0.425$\\
         GW191127\_050227 & GWTC-3 & 983 & 3.588$\cdot10^9$ & 58814.2 & $0.418$\\
         GW191129\_134029 & GWTC-3 & 856 & 5.496$\cdot10^7$ & 58816.6 & $0.436$\\
         GW191204\_110529 & GWTC-3 & 3380 & 3.436$\cdot10^9$ & 58821.5 & $0.449$\\
         GW191204\_171526 & GWTC-3 & 256 & 7.520$\cdot10^6$ & 58821.7 & $0.284$\\
         GW191215\_223052 & GWTC-3 & 586 & 4.535$\cdot10^8$ & 58832.9 & $0.299$\\
         GW191216\_213338 & GWTC-3 & 206 & 1.280$\cdot10^6$ & 58833.9 & $0.282$\\
         GW191219\_163120 & GWTC-3 & 2232 & 7.504$\cdot10^7$ & 58836.7 & $0.465$\\
         GW191222\_033537 & GWTC-3 & 2168 & 3.687$\cdot10^9$ & 58839.1 & $0.417$\\
         GW191230\_180458 & GWTC-3 & 1086 & 4.376$\cdot10^9$ & 58847.8 & $0.247$\\
         GW200105\_162426 & GWTC-3 & 7882 & 3.345$\cdot10^7$ & 58853.7 & $0.438$\\
         GW200112\_155838 & GWTC-3 & 3200 & 5.599$\cdot10^8$ & 58860.7 & $0.352$\\
         GW200115\_042309 & GWTC-3 & 512 & 3.792$\cdot10^6$ & 58863.2 & $0.329$\\
         GW200128\_022011 & GWTC-3 & 2415 & 5.729$\cdot10^9$ & 58876.1 & $0.357$\\
         GW200129\_065458 & GWTC-3 & 31 & 2.617$\cdot10^6$ & 58877.3 & $0.857$\\
         GW200202\_154313 & GWTC-3 & 150 & 2.155$\cdot10^6$ & 58881.7 & $0.872$\\
         GW200208\_130117 & GWTC-3 & 30 & 3.075$\cdot10^7$ & 58887.5 & $0.000$\\
         GW200208\_222617 & GWTC-3 & 2040 & 1.214$\cdot10^{10}$ & 58887.9 & $0.397$\\
         GW200209\_085452 & GWTC-3 & 877 & 2.325$\cdot10^9$ & 58888.4 & $0.664$\\
         GW200210\_092255 & GWTC-3 & 1387 & 1.595$\cdot10^8$ & 58889.4 & $0.555$\\
         GW200216\_220804 & GWTC-3 & 2924 & 1.113$\cdot10^{10}$ & 58895.9 & $0.750$\\
         GW200219\_094415 & GWTC-3 & 781 & 1.902$\cdot10^9$ & 58898.4 & $0.435$\\
         GW200220\_061928 & GWTC-3 & 4477 & 4.333$\cdot10^{10}$ & 58899.3 & $0.301$\\
         GW200220\_124850 & GWTC-3 & 3129 & 1.185$\cdot10^{10}$ & 58899.5 & $0.475$\\
         GW200224\_222234 & GWTC-3 & 42 & 1.947$\cdot10^7$ & 58903.9 & $0.143$\\
         GW200225\_060421 & GWTC-3 & 498 & 8.177$\cdot10^7$ & 58904.3 & $0.533$\\
         GW200302\_015811 & GWTC-3 & 6016 & 2.778$\cdot10^9$ & 58910.1 & $0.230$\\
         GW200306\_093714 & GWTC-3 & 3907 & 4.302$\cdot10^9$ & 58914.4 & $0.459$\\
         GW200311\_115853 & GWTC-3 & 35 & 5.799$\cdot10^6$ & 58919.5 & $1.000$\\
         GW200316\_215756 & GWTC-3 & 187 & 3.634$\cdot10^7$ & 58924.9 & $0.217$\\
\end{longtable}
\twocolumn
\bsp
\label{lastpage}
\end{document}